# Dynamical patterns of cattle trade movements

Paolo Bajardi[1,2], Alain Barrat[2,3], Fabrizio Natale[4], Lara Savini[5], Vittoria Colizza[6,7,8*]

[1] Computational Epidemiology Laboratory, Institute for Scientific Interchange (ISI), Torino, Italy

[2] Centre de Physique Théorique (Centre National de la Recherche Scientifique UMR 6207), Marseille, France

[3] Complex Networks & Systems Lagrange Laboratory, Institute for Scientific Interchange (ISI), Torino, Italy

[4] European Commission, Joint Research Center, Institute for the Protection and Security of the Citizen, Ispra, Italy

[5] Istituto G. Caporale, Teramo, Italy

[6] INSERM, U707, Paris, France

[7] UPMC Université Paris 06, Faculté de Médecine Pierre et Marie Curie, UMR S 707, Paris, France

[8] Institute for Scientific Interchange (ISI), Torino, Italy

[*] Corresponding author: vittoria.colizza@inserm.fr

## Abstract

Despite their importance for the spread of zoonotic diseases, our understanding of the dynamical aspects characterizing the movements of farmed animal populations remains limited as these systems are traditionally studied as static objects and through simplified approximations. By leveraging on the network science approach, here we are able for the first time to fully analyze the longitudinal dataset of Italian cattle movements that reports the mobility of individual animals among farms on a daily basis. The complexity and inter-relations between topology, function and dynamical nature of the system are characterized at different spatial and time resolutions, in order to uncover patterns and vulnerabilities fundamental for the definition of targeted prevention and control measures for zoonotic diseases. Results show how the stationarity of statistical distributions coexists with a strong and non-trivial evolutionary dynamics at the node and link levels, on all timescales. Traditional static views of the displacement network hide important patterns of structural changes affecting nodes' centrality and farms' spreading potential, thus limiting the efficiency of interventions based on partial longitudinal information. By fully taking into account the longitudinal dimension, we propose a novel definition of dynamical motifs that is able to uncover the presence of a temporal arrow describing the evolution of the system and the causality patterns of its displacements, shedding light on mechanisms that may play a crucial role in the definition of preventive actions.

# Introduction

Animal movements represent a crucial aspect for the trading and marketing of livestock, though they may offer an easy mean for rapid dissemination of zoonotic infectious diseases among animal holdings, with a spatial extent covering large geographical distances, as shown for example by the 2001 Foot-and-Mouth disease epidemic in the UK [1]. Animal diseases may compromise livestock welfare and reduce productivity, and may in addition represent a threat to human health, since the emergence of human diseases is dominated by zoonotic pathogens [2]. Disease management and control is thus very important in order to reduce such risks and prevent large economical losses [3]. This can be achieved for example through controlling animal movements and mixing, controlling entry to farm lots, quarantining animals, or imposing standstill periods that prevent further movements of animals off premises. To correctly evaluate such preventive and control measures, a detailed knowledge and regulation of animal movements is needed. A crucial step into addressing this issue has been taken in Europe by establishing the implementation of a digital framework for the identification and registration of bovine animals [4], and similar cattle identification and tracing systems have also been implemented in other countries [5]. Detailed data on the movement of individual cattle at the national level have thus become available that trace each bovine along its movements among premises on a daily basis.

Such monitoring efforts have led to a unique opportunity of studying animal movements in a detailed way, characterizing their behavior in time and space, and identifying patterns that may become relevant for the spread of a potential disease in the cattle population. A natural description of these systems is offered by the network representation in terms of nodes (the elements of the system, i.e., the premises in the cattle flow case) and links (the interactions among its elements, i.e., the cattle movements among premises) [6-11]. Much research has been done in the analysis of networked systems available from similar empirical datasets. The study of biological networks and transportation infrastructures, technological networks, human communication and mobility patterns [12-24] has unveiled the presence of unexpectedly similar properties, shared by these systems independently of their function, origin and scope. Besides the small world property, which consists in the co-existence of high local interconnectedness and small distances across any two nodes in the network compared to the system size [25], the components of such systems are found to be wired in a non-homogeneous way, with the number of connections per node showing very large fluctuations in contrast with the random Poissonian hypothesis [6]. The ubiquitous nature of this so-called scale-free property – found across natural, societal, and artificial systems – has spurred more than a decade of research aimed at characterizing and understanding complex systems drawn from different disciplines through the common paradigm of networks science [26].

The application of network approaches to veterinary medicine is however rather new. As reviewed recently by Dubé et al. [27] and by Martinez-López et al. [28], few papers have been published that analyze livestock movements by constructing the network of displacements and studying the relations between nodes with a systemic approach, thus going beyond the simple characterization of single-node properties (as e.g. the amount or frequency of displacements on and off single farms). The availability of datasets of cattle flows in several countries has opened the possibility to explore systematically and in great detail these contact structures that represent the key driver of disease spread through infected animals moving from farm to farm or through other livestock operations such as markets and dealers. Research in this context can be divided into descriptive analysis of the datasets in order to assess the implications for disease control [29-36], and epidemiological studies aimed at reproducing historical epidemics through models, estimating epidemiologically relevant parameters from the contact structures, and realistically modeling disease spreading based on the movements data [34,37-42]. However, while zoonotic disease modeling counts today a variety of different approaches (see e.g. the review of Ref. [42]), analogous developments are not observed in the characterization of cattle movements, with research efforts so far limited to a very basic understanding of the system [29-35,37,40,41,43].

Most of the veterinary studies are indeed based on static representations of cattle flows – where the temporal information of the displacements is collapsed into few successive snapshots of the datasets [30-35,40,41,43], or explored through the time series analysis of simple global quantities [29,31,32,35,40] – or focus on the results of spreading simulations based on the dynamical network and on its static counterparts [37]. In the analyses performed so far, results have shown a large heterogeneity in the connectivity patterns among premises, with probability distributions for the number of incoming and outgoing connections (in-degree and out-degree, respectively) characterized by broad tails [30-32,34,35]. While the majority of premises have a small number of connections through animal movements to other premises, this feature indicates the presence of a small but non-negligible fraction of premises that instead are recipients or senders of animal movements from/to a large number of holdings [6]. Such results are typically obtained from the investigation of a static network obtained by aggregating data on the full available time window [30,31,34], and few examples of structures extracted from shorter aggregation times (such as e.g. monthly and weekly networks) have been investigated [32,35], without, however, exploring in a systematic way the stability of this feature across time. Fits to the power-law behavior of the degree distributions yield values around 2.1 for both in- and out-degree for the 2002 UK cattle movement data obtained for a specific 4-weeks time window [35]. However these results cannot be easily compared to the values of 2.5–2.6 obtained for the total degree distribution of monthly and weekly snapshots of 2005 French data [32], and 2.1–2.2 obtained for the annual network of the 2005 French data [32] and 2007 Italian data [30], given that in-degree and out-degree are not considered separately. Broad distributions have also been found in the annual number of movement events to/from single premises, and in the annual number of animals displaced [30,34,35], along with asymmetries showing larger fluctuations in the quantities describing the annual incoming fluxes with respect to those measuring annual outgoing fluxes [34]. However the robustness of these properties at shorter timescales, comparable to the typical timescales of some zoonotic diseases, has not been assessed.

From the point of view of single animals, several studies have shown that the number of movements per bovine is typically very low, reaching at most 7-10 displacements [31,44], with the majority of them occurring on short distances, though a non-negligible fraction cover very long distances [31,35], thus highlighting the possibility for rapid dissemination across the country. Investigations into the dynamics of individual bovines displacements has however not been analyzed with the aim of uncovering the presence of specific paths, motifs or cycles that may be relevant in spreading a disease or in creating recurrent patterns favoring the virus propagation from one farm to another.

Overall, the dynamical information has only been partially considered, through time series analysis of global quantities, such as the number of premises involved in the flows or the total mass of movements [29,31,32,35,40], whereas much attention has focused on the role of different premises in the structure and flows management on the annual scale [30,31,45]. With the aim of assessing the spreading potential induced by the complex structures hidden in the data, much work has been dedicated to the analysis of the components of the network (giant component, weakly and strongly connected components, etc.) in order to estimate the upper bounds of the epidemic size [29,30,32,34,35,45], and to the ranking of nodes in terms of various measures of centrality defined *a priori*, such as degree, betweenness, and others [30,33,45], that in other systems were found to impact the behavior of dynamical processes taking place on top of them [11,46-58]. The aim is to compare the efficacy of prevention and control measures based on this information [30,32], though no assessment of the stability of these features in time is provided, thus affecting the applicability of the same measures in different points in time, due to the time evolution of the network. A recent example to overcome this limitation is provided by Ref. [36] that integrates flows dynamics and centrality measures to assess farms' vulnerabilities.

While network analysis has become increasingly important in the study of data describing livestock movements, a time lag is clearly observed between the developments of network analysis tools and the knowledge reached so far in the context of veterinary epidemiology. As described above, only basic network analysis has been considered, often disregarding the longitudinal dimension and focusing on single node properties of the cattle flow networks. Network science, on the other hand, offers nowadays a body of sophisticate and very advanced techniques and methodologies that are able to uncover higher order topological correlations, non-trivial correlations between topology and flows, backbone structures carrying the most relevant information contained in the system, motifs, recurrent patterns and communities, and other features [6-11]. Moreover, the recent availability of large-scale longitudinal datasets [19,20,22,24,59-64] has opened a new set of issues and challenges to deal with intrinsically dynamical systems (see e.g. [65-68]), and spurred an intense research activity aimed at the inclusion of dynamical aspects through a newly defined set of analysis tools and frameworks.

Similarly to cell-phone data [20,22], cattle flow data represents a rare example of high resolution dataset where fully identified displacements are described at the agent level with a daily resolution spanning one or several years, and are associated to multiple possible definitions of weight for the connections and to additional metadata describing e.g. the type and location of premises. By leveraging on the network approach, we are able for the first time to fully characterize the dynamical patterns of cattle trade flows, by investigating snapshots properties and assessing their stability across time and the role of the chosen aggregating time window, characterizing the rules describing the evolution dynamics of the individual agents movements and of the activity of the system on a larger scale, exploring stationary properties and identifying recurrent patterns where causality relations between displacements emerge. Going beyond the analyses performed so far on livestock movement datasets, we aim at unraveling the hidden complexity of these systems at the topological, functional and dynamical levels, in order to identify patterns and properties relevant for the identification of vulnerabilities of the system to an epidemic and thus build upon this information to develop preventive and control measures. In addition, our analysis uncovers important limitations of the approximations generally used in modeling approaches for zoonotic disease spreading. While the present work focuses on the dataset of the Italian cattle trade movements, the general approach formulated here can be directly applicable to the study of other livestock movements datasets to uncover differences and similarities across countries, and the impact of diverse preventive measures or tracing systems adopted at the national level.

The paper is organized as follows. After the description of the dataset under study, we characterize the system in terms of successive snapshots obtained from aggregating the data on different time windows. This allows us to study the emergence and robustness of network properties across time and the role of the timescale of aggregation, in view of its interplay with typical disease timescales. We then analyze the dynamical evolution of the network, both at the agent level and at the system level, and explore its impact on the structural backbone of the system and the efficacy of control measures against the spread of a disease. Finally, we introduce a novel definition of dynamical motifs for a time dependent evolving network, able to uncover causal recurrent paths in the bovine movements.

# Methods

## Data

Data on cattle trade movements were obtained from the Italian National Bovine database, which is administered by the Italian National Animal Identification and Registration Database [68]. The database details the movement of the entire Italian population of bovines among animal holdings, providing a comprehensive picture of where cattle have been kept and moved within the country. Each movement record reports the unique identifier of the animal, the codes of the holdings of origin and destination, and the date of the movement. Such tracking system allows us to easily reconstruct the path of each bovine and to build the corresponding overall network, minimizing the problems related to data accuracy that are found in other tracking systems that do not provide both origin and destination of the displacements [34,35]. Additional information was provided for the animal holdings, including the type of premises (i.e. fattening farm, dairy farm, pasture, slaughterhouse, assembly center, market, genetic material center, and other), and their georeferenced metadata in terms of the geographic coordinates of the centroids of the municipality where the premises were located.

Here we examine the records for the year 2007 [30]. A total of 4,946,201 bovines were tracked, counting for 7,177,825 recorded displacements of individual animals and 1,592,332 distinct batches movements. There were 173,139 active premises during the year (i.e. they either received a batch or moved it), of which 49.9% were fattening farms, 26.1% were dairy farms, 1.7% were pasture, 1.1% were slaughterhouses, 0.4% were assembly centers, 0.06% were markets, 0.04% were genetic material centers, and the remaining 20.7% were labeled as other premises. Active premises are located on almost the entire territory of the country, covering 96% of the Italian municipalities, though their distribution is not uniform – a single municipality can indeed contain a number of holdings varying from few units to hundreds. A total of 365 days of activity was recorded, from January 1st to December 31st of 2007, signaling that at least one displacement per day took place in the year under study. The dataset also contains information on the importation and exportation of cattle; these movements, representing less than 1% of the total number of movements in the database, were however excluded from the analysis as the focus of our study is on the full set of displacements within national boundaries. Table 1 summarizes some basic properties of the dataset.

## Construction of daily and aggregated networks

The system of cattle trade movements can be represented in terms of a network, similarly to other mobility datasets and transportation systems [13-15,18,19,21,22,69-72]. The simplest representation is obtained when nodes correspond to premises, and a directed edge is drawn between two nodes whenever a displacement of bovines occurs between the corresponding premises. Since data on cattle movements is provided on a daily basis by the original dataset, it is thus possible to construct 365 daily networks, each containing the activity of nodes and links for one day. It is also useful to construct static snapshots of the system by aggregating the observed activity over various time windows $\Delta t$. This static view partially looses the intrinsic dynamical nature of the system within the given time window, however it allows to study the static snapshots with the usual techniques of network theory [7-11,30-35,40,41,43]. Given a specific choice of $\Delta t$, we can construct $365/\Delta t$ such consecutive snapshots, corresponding to the time windows $[n\Delta t,(n+1)\Delta t]$, with $n$ going from 0 to $365/\Delta t - 1$. In addition to the intrinsic time resolution of the system, $\Delta t = 1$ day, we also consider time windows of $\Delta t = 7$ days, $\Delta t = 28$ days (we avoid aggregating over calendar months to avoid fluctuations due to the different duration of the months during one year), and $\Delta t = 365$ days. These choices give rise to 365 daily networks, 52 weekly networks, 13 monthly networks, and one annual network, respectively, the latter aggregating the whole activity reported in the dataset. While in the literature annual and monthly networks have been typically analyzed (with the exception of Ref. [32] that studied the weekly networks of French data), here we consider different values of $\Delta t$ in order to systematically explore the dynamical features of the networks on a variety of timescales, comparable to the timescales of different diseases of interest. Given that the aggregation on a time window $\Delta t$ is a commonly used approximation to describe an epidemic process occurring at timescales much slower than $\Delta t$, this study allows us to assess the role of the aggregation under changes of $\Delta t$, and verify whether similar conditions and properties are observed by changing that value.

As the networks are directed, each node $i$ is characterized by both its out-degree $k_{i,out}$ (i.e., the number of premises to which a movement is registered within the given time window) and in-degree $k_{i,in}$ (i.e., the number of premises from which the node receives an incoming flux of animals within $\Delta t$). For each snapshot, we consider only nodes with $k_{i,in} + k_{i,out} > 0$, defining them as *active* nodes since they correspond to premises that have registered at least one incoming or outgoing displacement during the aggregation time window. Moreover, the links of these networks can be weighted according to two distinct definitions, measuring either the number of cattle batches moved or the total number of animals moved [34]. More specifically, we denote by $w_{ij}^B$ the amount of cattle batches movements recorded within the given time window $\Delta t$ from the holding $i$ to the holding $j$. The weight $w_{ij}^A$ instead indicates the total number of bovines moved from $i$ to $j$ during $\Delta t$. The first quantity provides a binary information on a daily basis, and counts the number of movements occurring in the time window $\Delta t$ under consideration; the second measures the magnitude of the movements. The introduction of two different definitions of the weight is useful in order to explore whether there are any trivial correlations among the two quantities, and to assess the limits of the approximation that uses less detailed data such as the number of movement batches, which are usually more readily available than the detailed movements of animals at the individual level [37]. This would be very important in the framework of modeling approaches based on real data. By following the usual definition of strength of a node in a weighted network [14], we denote by $s_{i,in}^{B(A)} = \sum_j w_{ji}^{B(A)}$ and $s_{i,out}^{B(A)} = \sum_j w_{ij}^{B(A)}$ the in-strength and out-strength of node $i$, respectively, quantifying the total numbers of incoming and outgoing batch (*B*) and animal (*A*) movements of the corresponding premises during $\Delta t$.

## Dynamical properties

The dynamical nature of the dataset we consider allows us to go beyond the analysis of successive static snapshots. The dynamical aspects concern both the bovines that are moved along the displacement network, and the network's structure. In the next sections, we will start by using standard measures such as the study of the evolution of sizes of the snapshots, or the statistical analysis of the properties of the bovines displacements. We will also study how the properties of the network's elements fluctuate over time. Moreover, we will introduce specific new tools and methods specifically tailored towards highlighting the consequences of dynamical aspects of the displacement networks.

# Results and Discussion

## Daily and aggregated networks

We first focus on the analysis of the networks aggregated at different scales, as described in the previous section. The analysis of these various static snapshots gives access to a first characterization of the system under consideration, investigating both its structural and dynamical properties. This allows for the first time the comparison of the features obtained at different timescales, and, for each timescale, the possible emergence of properties that remain stable or change across time, as the activity captured in each snapshot may indeed vary from one snapshot to another. Even the very basic features of the network, such as e.g. the number of nodes (noted *N*) and of edges, depend both on the time of the year at which we observe the system and on the duration of the aggregation $\Delta t$ [32]. Table 2 summarizes the basic properties of the aggregated networks for the various $\Delta t$ values considered.

At the smallest possible aggregation scale, $\Delta t = 1$ day, the networks are small and very sparse, including an average number of nodes of the order of few thousands (to be compared to the total of about $10^5$ nodes active across the whole year), and they are typically composed of small disconnected components, similarly to what was observed in the UK cattle data [44]. For larger values of $\Delta t$, i.e. longer aggregation times, an increasing number of nodes and links are present in the networks, since more and more distinct displacement events are registered during the time window. The average number of nodes of the weekly and monthly networks increases of one order of magnitude with respect to the daily case, as observed in the weekly and monthly snapshots of the French cattle data [32]. This number does not show great variations in time for a given $\Delta t$, however it remains very small if compared to the full network, thus indicating the presence of strong changes in the activation of nodes from one month to the other. When aggregating over time windows of increasing duration, the networks not only increase in size but also become denser, with the number of active connections growing faster than the number of active nodes. The small disconnected components observed in the daily networks coalesce, leading to an increasingly larger giant component (i.e. the largest connected component of the network). Though snapshots up to monthly networks are small in size compared to the total number of active nodes observed during the year, their structure and interconnectivity allows for the creation of giant components spanning a large fraction of the aggregated networks (e.g. more than 70% for $\Delta t = 7$ days, see Figure S1 of the online Supporting Information (SI)), similarly to what was observed in the analysis of cattle movement data in other countries [32]. Starting from daily networks that may offer only limited propagation at the daily scale, a giant component emerges if aggregating on timescales $\Delta t \geq 7$ that indicates the existence of paths of propagation from one node to another at the system level.

Figures 1 to 3 report a set of statistical properties of the networks generated by aggregating the data on time windows of lengths $\Delta t = 1, 7, 28, 365$ days. Given that each $\Delta t$ value corresponds to a set of snapshots (except in the case of $\Delta t = 365$ days), for the sake of visualization we show in each plot the distribution of the quantity under study for one particular snapshot chosen as an example (red circles), overlaid to gray lines that indicate the behavior displayed by the other snapshots corresponding to the same $\Delta t$ (in the weekly and daily cases, given the large number of snapshots, we show a random subset). This allows us to monitor the variations over time signaling changes of the system's statistical properties, as a function of changes in $\Delta t$ and the time of observation. Interestingly, these distributions are superimposed for successive time snapshots at a fixed value of $\Delta t$, denoting a statistical stationarity of global distributions, describing the activity taking place at the microscopic level. This behavior, which is observed here for the first time for cattle movement data, is consistently present for all $\Delta t$ under study, and is similar to what was observed in other systems for which longitudinal data is available, such as e.g. the airline transportation system analyzed in Ref. [24].

Figure 1 displays the distributions of in- and out-degrees. The in-degree distributions are broad, with a behavior close to a power-law and a slope approximately equal to -2. This is in agreement with the results found for a specific month of the 2005 UK cattle data [35], and shows that this behavior is a common feature of the system in various countries and, moreover, is independent of $\Delta t$. The range of values of $k_{in}$ clearly increases with increasing values of $\Delta t$. Large fluctuations are observed also in the out-degree distributions, however the range of possible values of $k_{out}$ is systematically one order of magnitude smaller than the corresponding range observed for $k_{in}$, not only for the annual network [34] but for every timescale investigated. Results show a clear asymmetry in the receiving and sending activities of the animal holdings, which can be explained by the typical activity of specific premises types, such as slaughterhouses, assembly centers and also markets. Such premises are indeed responsible for assembling cattle trade fluxes for commercial purposes, thus receiving batches from a large number of premises, assembling them and moving larger fluxes to fewer premises.

Similar probability distributions can be computed for the weights as well, taking into account the two possible definitions. Figure S2 of the SI shows how the weights $w^B$ have by definition a sharp cutoff at their maximum value $\Delta t$, therefore limiting the range of possible values they can assume in the case of small $\Delta t$. On the other hand, the number $w_{ij}^A$ of animals displaced between farms $i$ and $j$ is characterized by a broad distribution even for the shortest time window $\Delta t = 1$ day [31]. This shows how cattle displacements are most often characterized by a small number of animals, but that movements of very large numbers are also observed with a non-negligible probability. Interestingly, the shapes of the distributions are almost not affected by changes in $\Delta t$, denoting underlying non-trivial mechanisms that make these statistical properties stable across integrations on diverse timescales.

Figure 2 shows the in- and out-strength distributions, according to the two definitions for the weights. A pattern very similar to the degree distributions is observed: the in-strength distributions are broad even at small $\Delta t$, while the out-strength distributions broaden significantly only as $\Delta t$ increases, especially for $s_{out}^B$, which indicates the total outflow of batches. The asymmetry discussed above is thus retained if we consider the total number of animals displaced in and out of premises. Besides looking at the overall behaviors of these quantities in terms of probability distributions, it is interesting to explore whether non-trivial correlations arise that relate the topology with the flows at the premises level, by considering the correlations between the strengths and degrees of nodes. Figure 3 shows the results obtained when the strengths are defined in terms of the weights $w^B$ and $w^A$, considering both the inflow and outflow dynamics. The behavior is linear for the in-strength, signaling an absence of correlation between the number of premises from which a specific holding receives batches and the number of batches or bovines received on each connection [14]. In the case of the out-strength we observe instead a slightly superlinear trend when $s_{out}^A$ is expressed as a function of the out-degree, showing that more active farms in terms of number of connections also tend to send more animals on each connection [14], explaining the asymmetry observed before in the variations of $s_{out}$ and $k_{out}$ with respect to $s_{in}$ and $k_{in}$.

For increasing time window lengths, the aggregated networks take into account more displacement events. Concerning the links and nodes present in a network at a given timescale $\Delta t$, this means that their weights, degrees and strengths are expected to increase when longer time windows are considered. Notably however, we do not observe a simple shift of the whole distributions towards larger values with a corresponding absence of small values: the distributions continue to be broad, spanning several orders of magnitude, but the most probable values remain very small. In the case of the degree distributions, this can be due to nodes that have very few connections for any time window, or to nodes that are active only very rarely. For the weights distributions, it shows that *on any timescale* there exists many links that are active only during few days, already indicating the presence of a non-trivial underlying dynamics that cannot be uncovered through the analysis of static snapshots only.

## System dynamics

The results of the previous subsection show how the microscopic dynamics of cattle movements is described by statistical properties that are found to be stationary, with a behavior that is qualitatively invariant with respect to changes in the timescale (whereas size and magnitude of fluctuations clearly depend on the time window $\Delta t$). Here and in the following subsections we aim at characterizing the underlying dynamics to uncover higher order correlations and relevant temporal aspects leading to the observed behavior.

The simplest dynamical information is given by the evolution of the sizes of the aggregated networks. The numbers of nodes and links follow consistent patterns (as shown in Figure S3) with both weekly cycles and clear seasonal properties that distinguish the livestock activity across the different seasons [29,31,32,35,40]. On a monthly time scale, it emerges that the summer activity is substantially lower than the activity registered during the rest of the year. The evolution of the daily snapshots sizes shows moreover how the overall movements decrease strongly during the weekends, leading to increasingly smaller and more fragmented networks that put obstacles to the propagation of a disease across the system.

Differently from human mobility data where the information is usually not provided at the individual level and is aggregated into flows that cannot be traced back to the individual's behavior [14,69], the cattle movement dataset provides detailed information at the individual level through tracking each single animal during its displacements. This allows two different levels of description of the dynamics: (i) the agent-centered point of view that considers the features of the animals' movements (similarly to what can be done for individuals based on anonymized phone cell data [22,23]); (ii) the network point of view that focuses instead on the system's behavior and is given by the evolution of the topology, and of the links' and nodes' properties from one time window to the next. These views provide complementary information for the characterization and understanding of the dataset.

## Agent-centered dynamics

Gaining insight from this point of view aims at characterizing the trajectories of each bovine, uncovering the possible presence of predictable patterns, similarities or large heterogeneities, in the perspective of understanding the potential for disease propagation across the system, through its agents.

As bovine displacements are subject to livestock commercial constraints, we expect that the resulting bovine mobility patterns will be different from human mobility patterns [19,22,23,69]. Indeed, the number of displacements of any single animal over one year is quite restricted [31,44], as shown in Figure S4, in particular if compared with human behavior. On average bovines experience 1.45 displacements during a year. Interestingly however, some animals perform more than 10 moves, which may potentially result in superspreader behaviors. Figure S4 also reports the distributions of geographical distances covered either in a single displacement (i.e., the geographical distance between the origin and destination farms), or following the trajectory of a single animal in one year. Despite a well-defined maximum at short distances, these distributions display rather broad tails corresponding to very long routes [31,35]. In addition, the distributions are robust against filtering on the weight $w^A$, indicating that very long routes are performed by both small and long batches. The possibility of such long displacements should be taken carefully into account when dealing with spreading of diseases, as they could result in epidemics rapidly reaching geographically very distant parts of the network.

Another interesting issue concerns the time interval between two consecutive displacements of an agent, corresponding to the period that a given bovine spends in the same holding [31]. This time interval may represent the time of exposure of the animal to a potential outbreak taking place in the holding, or the time during which it could spread the disease to other animals if infected. Since the different types of farms have different roles in the bovines trade, the global distribution shown in Figure 4A is a convolution of several different behaviors. In particular, the two peaks at 3 and 6 months correspond to pasture and fattening farms, respectively, as shown by the other panels of the Figure that disaggregate the results by premises type. Except for the markets, in which bovines spend only few days, the distributions of these time intervals are broad for each farm type, with different slopes. This points out the large variety of possible timescales characterizing the time during which an animal stays in a given premises, indicating that homogeneous assumptions on the length of stay of an animal at a given holding do not provide an accurate description of reality. The broadness of these distributions should therefore be taken cautiously into account in the modeling approaches.

## Network microscopic dynamics

By comparing the results obtained for the weekly and monthly networks with those corresponding to the whole dataset, it is clear that a strong dynamical activity shapes the evolution of the system on both global and local scales. As an example, we show in Figure 5 a visualization of a subgraph for three consecutive monthly networks. The subgraph is constructed by selecting a particular seed node (the same for all three networks) and by considering all nodes at distance $\leq 3$ from the seed (where the distance is defined by the number of links traversed on the shortest path connecting the two nodes). Nodes keep their position in the visualization if they are active over multiple snapshots. The figure highlights how the structure of the neighborhood of a given node obtained at consecutive time snapshots can widely differ: even highly connected nodes in one snapshot can disappear from the neighborhood of the given node in the next snapshot, and hubs suddenly appear that were absent from the previous snapshot.

**Activity timescales.** Similarly to the dynamics of single animals, the network dynamics can be first characterized by the distributions of the activity and inactivity periods of nodes and links [24]. These periods are defined, for a given timescale $\Delta t$, as the number of consecutive time steps in which a node, or a link, is active (or not active, respectively). In the case of time windows of $\Delta t = 1$ day, we remove the weekends from the dataset as they are characterized by a much smaller activity, and consider a node or a link to be continuously active if it is present in the snapshots of a given Friday and of the Monday of the following week. The corresponding distributions are shown in Figure 6 for $\Delta t = 1$ and 7 days. As seen also in the dynamics of the air transportation network [24], most nodes and links turn out to be continuously active or inactive for only very short periods. The distributions of activity periods $\tau$ are rather narrow in the case of daily networks, and can be fitted by power-laws with exponent smaller than -4: most nodes and links are active only for one day at a time, and only very few are continuously active for more than a few days. The distributions become significantly broader when considering weekly networks, where power-laws with exponents close to -3 emerge. The difference observed by comparing $\Delta t = 1$ and 7 days can be easily explained by the integration over multiple days in the case of $\Delta t = 7$: being active in two consecutive such networks is a less stringent condition than being active each day of two successive weeks. The inactivity periods $\Delta \tau$ are characterized by much broader distributions extending on all possible timescales, signaling that a node (or a link) may become active at a given point in time without then participating to the dynamics for a long time interval. From the point of view of control policies, such long inactivity periods would help in limiting the spread through self-isolation of premises.

Given that the activities of nodes and links of the displacement network occur at both short and long timescales, here we aim at characterizing the mechanisms behind the appearance and disappearance of links in the system, and we focus on the weights $w^A$ that measure the number of animals displaced along each link. As proposed in [24] we evaluate in particular the fraction of appearing $f^a$ and disappearing $f^d$ links, as a function of their weight, in order to uncover a possible correlation between a link's stability and the number of displaced animals along that link. More precisely, if $E(w|t)$ is the number of links with weight $w = w^A$ at time $t$ and $E^a(w|t)$ is the number of such links that were not active at the previous time (and thus appeared at time $t$), the fraction of appearing links is $f^a(w) = E^a(w|t)/E(w|t)$. An analogous procedure leads to the definition of $f^d$ by considering the links of weights $w$ active at time $t-1$ but no longer active at time $t$. The quantities $f^a(w)$ and $f^d(w)$ are shown in Figure 7 for daily, weekly and monthly networks. We observe that $f^a(w)$ and $f^d(w)$ have an almost identical behavior, though dependent on the timescale $\Delta t$. In Figure S5 the results are disaggregated by premises type for the origin (or for the destination) of each considered link, showing that the behavior observed in Figure 7 results from a convolution of trends that are quantitatively different but qualitatively similar for all farm types. In all cases, links with small or large displacements of animals are both very unstable, whereas the most stable links are those with an intermediate weight. While till now the system of bovine movements showed properties that are very similar to those found in the analysis of human mobility by air travel, this result instead strongly differs from the positive correlations of links' stability and weight found in the airline transportation network [24]. In the airline system this is due to the fact that links with large weights correspond to busy routes that are economically convenient carrying a large fraction of the traffic and thus well established. Different commercial driving forces characterize the cattle trade flows and, in addition, premises have limited receiving capacities, constrained by the limited size of the space hosting the cattle for a widely varying number of days (see the results in Figure 4). Since a large weight corresponds to a transport of a large number of animals, it is rather unlikely that two (or more) very large such events occur on the same connection in rapid succession, as this may correspond to a large increase in the population at the premises, if no animals are moved away. Differently from this process in which bovines stay at the arrival node after displacement, airline passengers either connect through an airport or leave the airport to reach their final destination, without thus increasing the population at the mobility node itself. The result is that large displacements are very stable in the airline case, whereas heavily fluctuate in the bovine case. The lack of possible identification of stable connections over time carrying large weights (and thus having a large spreading potential) seems to indicate the absence of a robust pattern of movements in the system that could be easily targeted by intervention measures aimed at controlling and containing the spread of a disease. This aspect will be explored in further detail in the next subsection when evaluating the evolution dynamics of the network backbone.

As expected, the minimum values of $f^a(w)$ and $f^d(w)$ are very close to 1 when considering the daily networks, meaning that more than 80% of the links present at a given day will disappear the day after (and similarly for the appearance of links). At such timescale the full dynamical nature of the network emerges. More stable structures are instead detected at larger aggregation times, when weekly and monthly networks are considered.

**Fluctuations of nodes and links properties.** In addition to characterizing the dynamics with which nodes and links can switch on and off their activity, here we study the evolution of nodes' and links' properties once they are active. In particular, the evolution in time of a link's weight $w_{ij}(t)$ is characterized by its growth rate $r_{ij}(t) = \log \frac{w_{ij}(t+1)}{w_{ij}(t)}$ whose distribution is shown (for the weights $w^A$) in Figure 8 for the various time windows under study. The distributions are stationary, with exponentially decaying tails, as found for the airports network [24] and in studies of firm growth [72]. This corresponds to a weights' evolution from one month to the next of the form $w_{ij}(t+1) = w_{ij}(t)(1+\eta_{ij})$ where the multiplicative noise $\eta = e^r - 1$ is a random variable whose distribution is broad and does not depend on time, indicating that most of the weights increments are small but that sudden and large variations of the weights can be observed with a small but non negligible probability. The highly dynamical nature of the network, characterized by large instabilities and timescales describing the appearance and disappearance of nodes and links, is expected to have a strong impact on the nodes' properties as well. For instance, a node with many connections on a certain day may be much less connected the next day [73]. The stationarity of the distributions obtained from the analysis of static aggregated networks does not imply the stationarity of the properties of each given node; the set of nodes in the tail of the distribution may for instance differ from one snapshot to another. If we focus on properties of centrality of the nodes, which are often used to identify and target the elements of the system for isolation and quarantine aiming at prevention and control of an epidemic spreading on the network, large fluctuations in these values point to the strong limitations of such measures. In order to investigate this, we show the variations of a node's property for all snapshots considered, depending on the timescale $\Delta t$ under study. Figure 9 shows the median and the 95% confidence interval of all values of the out-strength $s_{out}^A$ that each node assumes when active, for different time window lengths. Very large fluctuations are observed, with most nodes showing variations over more than 2 decades, signaling that this property lacks stationarity at the node level. Some nodes with very high strength seem to have no fluctuations, but they appear in fact only once in the dataset. Similar results are obtained when considering other possible measures of node centrality, such as the in-strength or the in- and out-degree (not shown). Given that these quantities are proxy measures for the centrality of nodes, such findings strongly undermine the efficacy of traditional measures for epidemic control that do not take into account the large variations in time of the role of the premises with respect to the flows of the system.

## Evolution of network backbone

The results of the previous section show how the system is characterized by large fluctuations and strong topology and traffic variations on all spatial and temporal scales. The overall picture is thus one of a network whose structure changes very strongly from one snapshot to the next, not only at the global level, but also at the node neighborhood and node levels, inducing very strong centrality fluctuations. Notably, these centrality fluctuations are observed for all premises and geographical positions. A natural question therefore arises concerning the possible existence of a backbone of nodes and connections, carrying the relevant topological and dynamical information of the system, and of its temporal stability. The observed strong fluctuations may indeed be related to less meaningful connections of the network, and may thus be compatible with a stationary backbone.

A first attempt at defining a global backbone over time consists in considering the intersection of successive aggregated networks. If a considerable fraction of the system is stable across time, the intersection will be quite large and will identify the subset of premises and flows that have a predominant role in the dynamics. At the monthly scale, less than 4% of the links are common to the 13 corresponding networks. Moreover, the corresponding weights are not particularly stable and show a growth rate distribution comparable to the one obtained for the original networks. More advanced filtering techniques can be used to extract a statistically significant subgraph that carries a significant part of the traffic. In particular, it is possible to retain only the links with weights larger than a certain threshold, i.e., the ones with most traffic. However, when the system is characterized by large fluctuations of weights, and more in detail by a large heterogeneity of weights around a given node (as is the case here, not shown), global thresholding can lead to misleading dismissal of locally very important links [74]. For this reason, in addition to the thresholding method, we consider the disparity filter method that was introduced in Ref. [74]. For each node, it consists in identifying the links that should be preserved in the network. To this aim, one considers the null hypothesis of a random assignment of the normalized weights $p_{ij}^{out} = w_{ij}/s_{i,out}$ and $p_{ij}^{in} = w_{ij}/s_{j,in}$ (as links are directed, we consider two normalized weights definitions), and computes for each link the probability $\alpha_{ij}^{in(out)}$ that its normalized weights are compatible with this hypothesis. These probabilities are given by $\alpha_{ij}^{in} = (1 - p_{ij}^{in})^{k_{j,in}-1}$ and $\alpha_{ij}^{out} = (1 - p_{ij}^{out})^{k_{i,out}-1}$ [66], and the backbone is given by the links which satisfy at least one of the conditions $\alpha_{ij}^{in} < \alpha$ or $\alpha_{ij}^{out} < \alpha$, where $\alpha$ is a parameter that can be tuned in order to change the significance level of the filtering. For networks with uncorrelated weights, the disparity filtering procedure is equivalent to the global thresholding procedure, pruning the links with a weight smaller than $\langle w \rangle \cdot \ln(1/\alpha)$, where $\langle w \rangle$ is the average weight in the network. We have considered both filtering procedures for the aggregated networks with $\Delta t = 28$ days, and constructed for each network the corresponding backbones for various significance levels. In order to assess how the network backbone change in time, we have then computed the overlap between the $13 \times 12/2$ pairs of backbones, at a given significance level, where the overlap of two networks with respective sets of edges $E_1$ and $E_2$ is defined by $|E_1 \cap E_2|/|E_1 \cup E_2|$. Figure 10 displays the distributions of the growth rates of the links' weights $w^A$ in the various backbones, compared with the corresponding distribution in the whole network, together with the overlaps of the monthly backbones in color-coded matrices. The overlap between backbones of successive monthly networks is substantial but not large, approximately ranging from 25 to 30%. If we assume that the two successive backbones have the same size, an overlap of 25-30% would correspond approximately to an intersection of 40-50% between the two systems. While about half of the system is retained from one snapshot to the following, this value becomes rapidly smaller when moving away from the diagonal, i.e., as the corresponding networks are further apart in time. This shows that the *memory* of the most significant links in a given month rapidly fades away in the successive months, and that evaluating the importance of a link based on previous evidence could thus be misleading.

## Percolation

The very short memory of the backbone structure leads us to the study of how the dynamical aspects impacts the percolation properties of the network of displacements. Percolation has long been used in the analysis of complex networks [75,76], and results have shown that many real-world network structures typically retain their integrity, in terms of global connectedness, when nodes or links are removed in a random fashion, while they are very fragile with respect to targeted attacks. In this respect, percolation analysis has become a tool to investigate the structure of networks, by studying how the size of the largest connected component evolves when nodes are removed according to different procedures [75-79]. The size of the giant component not only is a measure of the resilience of the structural properties of the network under study, but it also quantifies the extent to which an epidemic could possibly spread in the system. Identifying ways to reduce this size, by removing particular nodes, is equivalent to finding efficient intervention and control strategies in the framework of disease spreading, aiming at breaking down the network in small pieces in order to prevent the disease from invading the system.

Let us consider for instance that an outbreak starts at a certain date. It is then possible to sort the nodes of the aggregated network of the corresponding time window by their degree, strength, or other centrality measures, and to try to contain the disease spread by isolating the most central nodes, reducing drastically the size of the largest connected component through the isolation of only a few percents of the nodes [75,77,78]. A lot of work has been done in this direction for the analysis of the fragmentation of the network of livestock movements when nodes are chosen according to different centrality measures [29,30,32-35,45], and no information on disease spreading is considered, as instead was done in Ref. [36]. However, these studies have neglected the dynamical nature of the system, focusing on specific snapshots only, and assuming to be able to access all the relevant information of the system at any given point in time, e.g. during an epidemic emergency. Since we showed so far how the underlying topology and flows strongly fluctuate at all levels, here we want to study instead the situation in which we have limited information on the system gathered from its activity on the last time window under study, and we want to apply isolation and quarantine measures to the following snapshot. The ranking of nodes according to a given centrality measure (corresponding to their spreading potential) computed on a certain time window may indeed loose its relevance when applied at successive times. Given these intrinsic dynamical features, we aim here at assessing the impact of a removal strategy on consecutive snapshots (thus the snapshots characterized by the highest values of the overlap), once the strategy is defined on the basis of the available information on one snapshot only, and is not updated according to the successive network evolution.

We investigate this aspect by measuring the effect of the successive removal of nodes by decreasing degree in consecutive snapshots of $\Delta t = 28$ days. More in detail, by focusing on a given snapshot for $\Delta t = 28$ days (the third snapshot of the year, chosen as an illustrative example), we fix the order of nodes to be removed in a degree-decreasing fashion. Then, we assess the impact of the removal of nodes ordered in such way on this snapshot and on the following one. This means that for the successive snapshot we are not re-evaluating the centrality of each node (as measured here by the degree) but we use the information computed on the previous time snapshot. This procedure is tested on the full network and on the corresponding backbone, calculated at two different significance levels. Figure 11 shows the results in terms of the relative size of the giant component as a function of the fraction of nodes removed. As expected, the removal of nodes is very efficient if the order of nodes to be removed is calculated on that snapshot [29,30,32-35,45], whereas such ordering is not able to destroy the network at the successive time window, leading to a size of the giant component that decreases very slowly, and maintains a fraction of more than 20% of the system still intact and connected after the list of nodes is exhausted. Even though a large number of nodes is removed from the system, the effectiveness of such isolation procedure is strongly limited by that fact that the premises' properties have dramatically changed. Many of the active premises have appeared/disappeared from one snapshot to the other, and the ones that remained have strongly changed their interaction pattern. In such situations, intervention and control strategies devised using the information from static aggregated networks, or more generally from data from past mobility patterns, can thus result to be very inefficient.

## Dynamical motifs

After observing the large fluctuations and the fast dynamics characterizing the system at all timescales, the last analysis we present in this subsection aims at going even further in the understanding of the system flows by exploring the possible signatures of a temporal ordering of the bovine displacements and the presence of recurrent paths.

One of the main consequences of the temporal evolution of the network resides in the causality constraints it induces. For instance, a spreading phenomenon can propagate on a path $ijk$ (i.e. from $i$ to $j$ to $k$) only if the link $ij$ is present before the link $jk$. The search in networks of the abundance of particular paths or motifs [80] should then be complemented by causality requirements and approaches that are able to incorporate the longitudinal dimension [64,65,81,82]. This becomes particularly relevant if the flows form cycles or paths that allow the re-infection of some premises, given an appropriate interplay of the disease and movement timescales. From the point of view of quarantine and similar control strategies, this would represent an important phenomenon to take into account when establishing the identification of premises to isolate, or the durations of disease surveillance at those locations.

Figure 12 presents an example of causal motifs: for instance, the repetition of the sequence of a link $ij$ followed in the next snapshot by a link $jk$ could imply a cause-effect relationship between these two links. Here we introduce a new measure to define causal motifs and restrict our analysis on the shortest possible timescale, i.e. the intrinsic timescale of the system $\Delta t = 1$ day. We collect, for each path length $l$, the motifs given by a list of links $i_0i_1, i_1i_2, ..., i_{l-1}i_l$ such that $i_0i_1$ is present at a certain snapshot $t_0$, $i_1i_2$ at snapshot $t_0 + 1, ...$, and finally $i_{l-1}i_l$ at $t_0 + l - 1$. The duration of the path is therefore equal to its length, and each path corresponds to a possible propagation that respects causal constraints. Each motif can occur for several values of the starting time $t_0$, and motifs of a given length can be ranked according to their number of occurrences. It is worth remarking that the above definition does not focus on the shape of the motifs since only temporally connected chain-like motifs are considered, and the recurrence is sought at the microscopic level counting the number of appearance of a certain link sequence. Figure 12 shows the corresponding frequency-rank plots, as well as the fraction of motifs that are repeated more than once for each length. Motifs are found up to length $l = 8$, and both the absolute number of motifs and the fraction of recurring motifs strongly decrease for increasing lengths. Since the number of times a link is present in the daily networks is broadly distributed (this number is given by the weight $w^B$ in the globally aggregated network over the whole year), pure statistical effects could be responsible for the abundance of specific patterns. For instance, if both links *ij* and *jk* are present all the time, then the causal path *ijk* will be very frequent. We therefore compare in Figure 12 the results obtained in the real data with different null models. The first null model (*random ordered*) is constructed by randomly shuffling the order of the daily aggregated networks: in this way, the structure of each daily network is kept, but the temporal correlations are lost. The second null model (*temporal mixed edges*) shuffles randomly the days in which each edge is active, independently from one edge to the next. The resulting daily aggregated networks have therefore randomized structures. Finally, we construct also a third null model by reshuffling the edges in each daily network as described in Ref. [83] (*time ordered and reshuffled networks*): we recall that this procedure consists in taking at random pairs of links *ij* and *lm* involving 4 distinct nodes, and rewiring them as e.g. *im* and *jl*. This procedure preserves both the in- and out-degree for each node, but destroys correlations. Figure 12 shows that the two first null models lead to similar results: a much smaller number of motifs is observed, and a smaller fraction of these motifs are found more than once. At lengths smaller than 5 however, this fraction is non negligible, showing that purely statistical effects due to the frequent presence of some links account for a part of the motifs presence and repetition. When both time ordering and network topology are reshuffled, motifs essentially disappear.

Since the network under study is directed, it is interesting to note that a causal sequence of links ($i_n i_{n+1}$ at a certain snapshot $t_n$, followed by $i_{n+1} i_{n+2}$ at snapshot $t_n + 1$) is not a valid causal path if it happens in the reverse order ($i_{n+1} i_{n+2}$ followed by $i_n i_{n+1}$). We therefore consider in Figure 12 also the sequence of 365 daily aggregated networks, seen in the reverse temporal order. Strikingly, the number of motifs is much smaller than for the true temporal sequence, and the fraction of repeated motifs is close to the case of a random temporal ordering. This indicates the presence of an intrinsic *time arrow* in the dataset, and provides a general method for investigating this aspect in dynamically directed networks. To our knowledge, this is indeed the first time that an intrinsic arrow of time has been explicitly detected in a temporal network. In Figure 12 we also show the number of motifs passing through a farm for different farm types. In order to take into account the relative abundance of the different farm types, we compare the results with a null model where the labels describing the farm types are reshuffled. We notice that some premises types (such as assembly centers or markets) are much more prone to be part of causal motifs than what would be expected for a random labeling of the premises. Our definition of causal motifs is therefore able to characterize the behavior of premises by identifying those types of premises that, as expected, show highly recurrent flow-in/flow-out patterns at such short timescale. The present analysis can also be extended by considering longer latency times for the occurrence of specific causal paths in the network, by considering sequences of links $ij$ at time $t$ and $jl$ at time $t + t'$, relaxing the previous condition on the separation of times between the occurrences of successive links in the motifs. The flexibility of this approach thus allows the tuning of the analysis to the relevant timescales of the spreading process under study, with a variable latency time $t'$ that corresponds to the time during which a node can be considered as continuously infectious. On the other hand, exploring different values of $t'$ allows us to explore possible scenarios of interventions through quarantine measures and isolation of premises of different duration, and assess their efficacy when simulating an epidemic process in the system.

# Conclusions

Empirical datasets characterizing cattle displacements are increasingly becoming available thanks to monitoring and tracking systems put in place in many countries, after recognizing the fundamental importance that movements have in disseminating an epidemic from one farm to another, with the potential of leading to national emergencies. By leveraging on the approaches and techniques of network science, in this paper we have presented a full analysis of the dynamical system of cattle movements, going beyond static and simple approximations and taking fully into account the temporal dimension of the dataset, using the Italian data of 2007 as a prototypical example. Starting from detailed data at the individual level at a daily resolution and covering a whole year, we have constructed aggregated networks on different timescales to characterize the system's behavior on a variety of timescales typical of different diseases, exposing the coexistence of stationary statistical distributions and strong microscopic dynamics at all time and spatial scales. We have shown how this dynamics affects not only global quantities (such as the number of connected nodes), but also the nodes' and links' properties at a very local level, and in relation with the rest of the system. In particular, the centrality of a node fluctuates strongly in time, thus preventing a straightforward static assessment of the spreading potential of premises that could be used for the definition of prevention and control measures. Longer historical data may be of help in assessing the role of specific premises at high risk of flow-in/flow-out situations. The network's dynamics also hinders the definition of a stationary backbone for the system structure and function, as a subset of the most important links (and weights) that are stable over time. We found indeed that the nodes and links forming the backbone strongly vary depending on the time window considered, and that the memory of the backbone rapidly fades away from one snapshot to the successive ones. This has important implications for the dynamical phenomena occurring on the system. Evaluating the information available at a given time step, to devise containment strategies against an epidemic spreading on the system, would indeed lead to inefficient measures if applied at other times. Finally, we have put forward a definition of dynamical motifs, formed by sequences of links that allow causal propagation, and illustrated how this definition can unveil the existence of an intrinsic time arrow in the dataset. The number of motifs of various lengths is indeed strongly different in the real dataset and in a time-reversed version; moreover such definition can be easily extended to focus on a variety of timescales of interest for the study of the disease spreading.

This study opens the road to future work in several directions. First of all, it would be interesting to explore such full characterization of the bovine movements dynamics also in other datasets corresponding to other countries, with the aim of uncovering similarities or differences, and assess how these may depend on different livestock market strategies and dynamics, or to the implementation of specific prevention measures. Also the effect of the introduction of new measures and regulations for the bovine movements and market could be possible with the analysis introduced here. Finally, the main result of our work highlights the non-trivial dynamical properties that prevent the study of the system from a stationary or quasi-stationary point of view and that have a strong impact on the dynamical processes that take place on this network. This opens a new challenge in the study of epidemic processes on dynamical substrates that continually evolve in time, with problems arising from the lack of stationarity, the interaction between multiple timescales, the strong dependence on the initial conditions, the presence of a non-reversible time direction, and others. Furthering our understanding of such systems through sophisticate tools of analysis and exploration of scenarios by modeling would be crucial to evaluate the behavior of such real-world systems under a disease emergency and help identify possible responses to minimize the epidemic impacts.

## Tables

**Table 1**. Cattle trade movements: Data from the Italian National Bovine database for the year 2007.

| Property | Value |
|---|---|
| Number of bovines | 4,946,201 |
| Number of animal movements | 7,177,825 |
| Number of batch movements | 1,592,332 |
| Average batch size | 4.5 |
| Days of activity | 365 [Jan 1 – Dec 31] |
| Number of active farms | 173,139 |
| Average number of days of activity per farm | 10.3 |
| Number of municipalities with active farms | 7780 (96% of the Italian municipalities) |
| Average number of active farms per municipality | 22 |

**Table 2.** Summary of the main features of the mobility networks obtained by aggregating the data over a time window $\Delta t$

| Aggregating time window | Variable | Average | Variance $\sigma$ | [min,max] |
|---|---|---|---|---|
| $\Delta t = 1$ day (365 networks) | # of nodes | 4.9 x $10^3$ | 3 x $10^3$ | [85, 1.1 x $10^4$] |
| | # of links | 4.2 x $10^3$ | 2.8 x $10^3$ | [49, $10^4$] |
| | $k_{in}$ | 0.9 | 6.2 | [0, 683] |
| | $k_{out}$ | 0.9 | 0.8 | [0, 178] |
| | $w_{ij}^B$ | 1 | 0 | [1, 1] |
| | $w_{ij}^A$ | 3.6 | 10.4 | [1, 2039] |
| $\Delta t = 7$ days (52 networks) | # of nodes | 2.6 x $10^4$ | 2.8 x $10^3$ | [1.5 x $10^4$, 2.9 x $10^4$] |
| | # of links | 2.8 x $10^4$ | 3.4 x $10^3$ | [1.5 x $10^3$, 3.2 x $10^3$] |
| | $k_{in}$ | 1.1 | 11.9 | [0, 1595] |
| | $k_{out}$ | 1.1 | 1.1 | [0, 178] |
| | $w_{ij}^B$ | 1.05 | 0.3 | [1, 2039] |
| | $w_{ij}^A$ | 3.8 | 11.9 | [1, 7] |
| $\Delta t = 28$ days (13 networks) | # of nodes | 6.4 x $10^4$ | 3.8 x $10^3$ | [5.6 x $10^4$, 6.9 x $10^4$] |
| | # of links | 9 x $10^4$ | 6.5 x $10^3$ | [7.8 x $10^4$, 9.9 x $10^4$] |
| | $k_{in}$ | 1.4 | 22.9 | [0, 4154] |
| | $k_{out}$ | 1.4 | 1.9 | [0, 219] |
| | $w_{ij}^B$ | 1.3 | 0.8 | [1, 25] |
| | $w_{ij}^A$ | 4.8 | 18.1 | [1, 2039] |
| $\Delta t = 365$ days (1 network) | # of nodes | 1.7 x $10^5$ | - | - |
| | # of links | 5.77 x $10^5$ | - | - |
| | $k_{in}$ | 3.3 | 59.5 | [0, 13186] |
| | $k_{out}$ | 3.3 | 7.0 | [0, 649] |
| | $w_{ij}^B$ | 2.7 | 5 | [1, 250] |
| | $w_{ij}^A$ | 9.8 | 65 | [1, 10845] |

# Figures

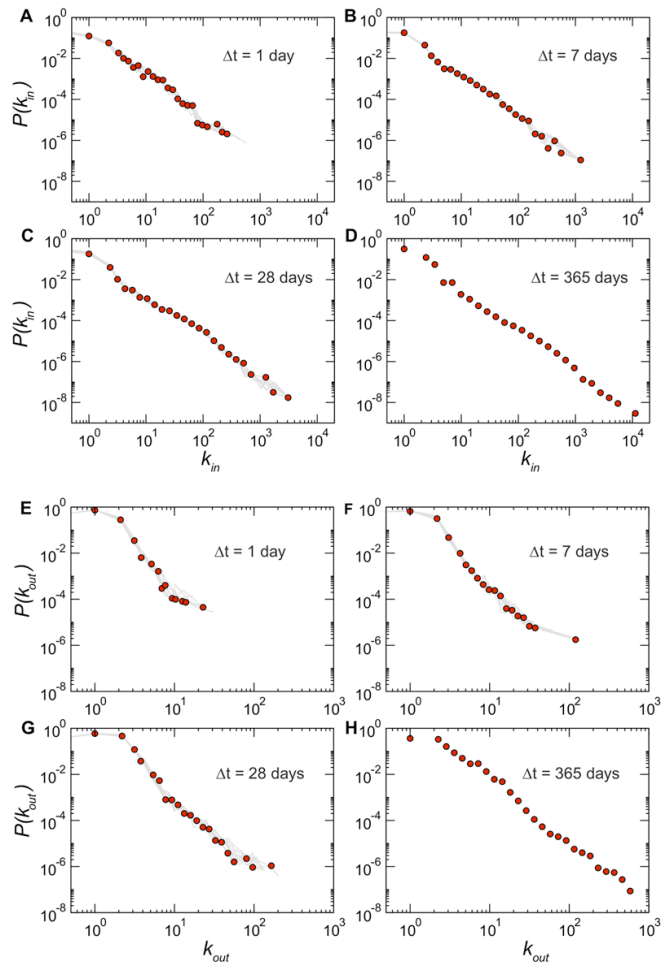

**Figure 1**. **Degree distributions for networks aggregated on different timescales** $\Delta t$. Since a single value of $\Delta t$ (for $\Delta t < 365$ days) yields multiple snapshots, each panel shows one distribution obtained for a given snapshot (circles) superimposed to a subset of the distributions obtained for the other snapshots at the same value of $\Delta t$ (grey lines). Panels A to D report the distributions of the in-degree $k_{in}$, that show very large fluctuations and a power-law like behavior with exponent close to $-2$ in all cases. Panels E to H present the distributions of the out-degree $k_{out}$, characterized by a cut-off that strongly depends on the length of the aggregating time window.

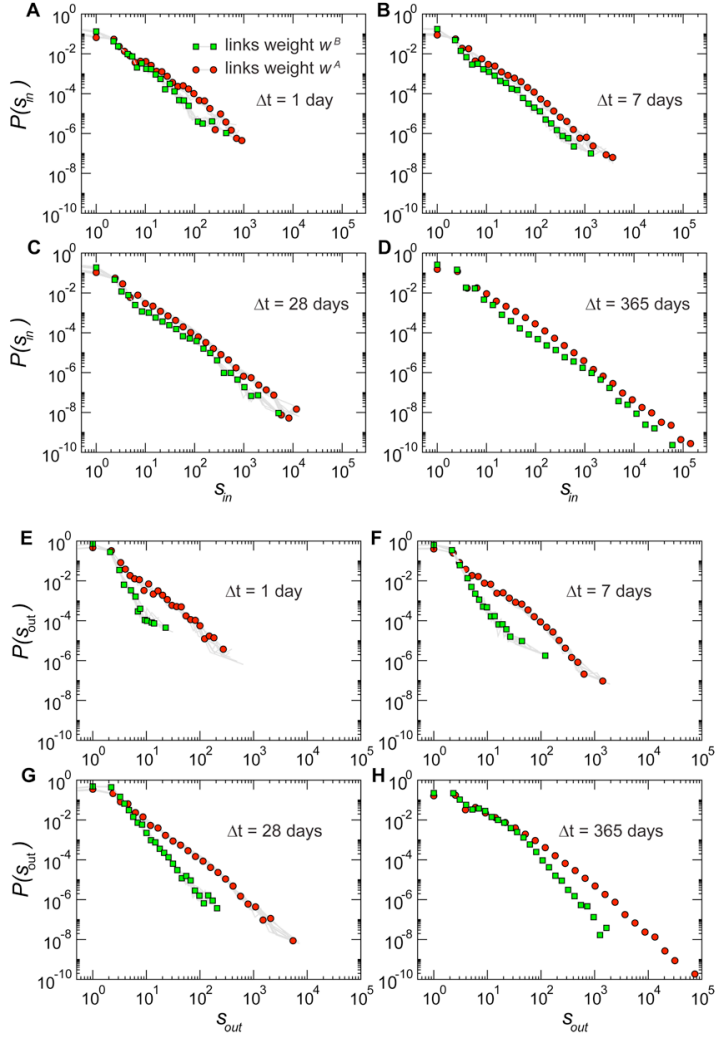

**Figure 2. Strength distributions for networks aggregated on different timescales $\Delta t$.** Panels A to D report the distributions of the in-strength $s_{in}$. Interestingly, the definition used to weight the links does not affect the distribution of the incoming traffic: the distributions $P(s_{in}^A)$ and $P(s_{in}^B)$ are very close. Panels E to H present the distributions of the out-strength $s_{out}$, whose behavior instead depends strongly on the type of weight considered. Broader tails are observed when considering the total number of animals displaced out of a given holding. The same representation of Figure 1 is adopted, with symbols representing the result of a particular snapshot, and grey lines the results obtained for a subset of the other snapshots.

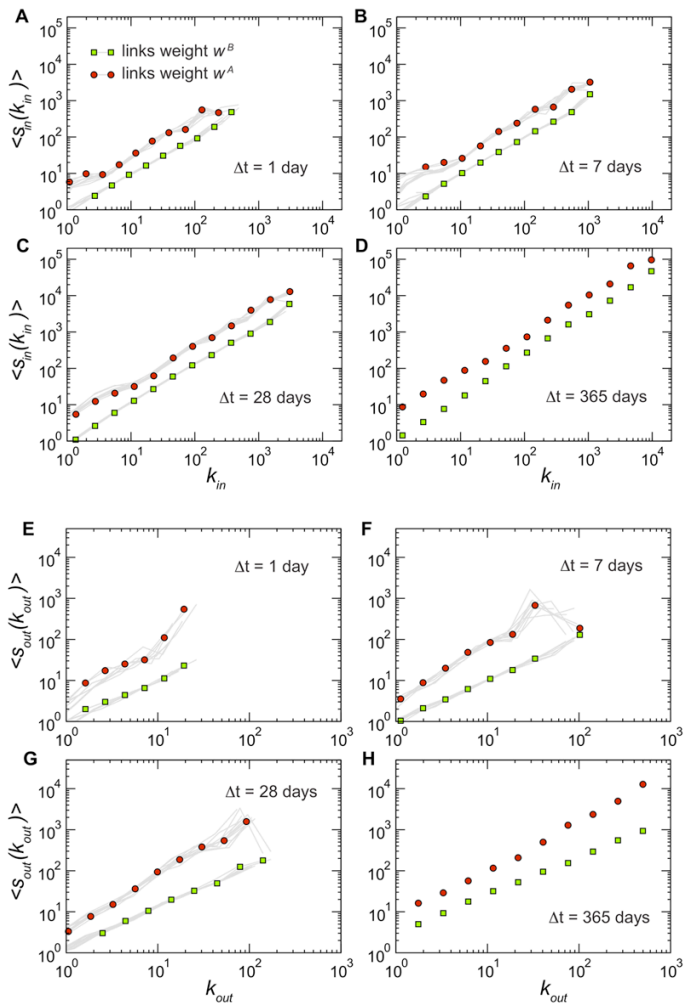

**Figure 3. Relation between the number of bovine traffic movements of a holding and its number of connections for different values of $\Delta t$.** Panels A to D report the average in-strength of nodes with a given value of in-degree, whereas panels E to H present the average out-strength of nodes with given out-degree. The same representation of Figure 1 is adopted, with symbols representing the result of a particular snapshot, and grey lines the results obtained for a subset of the other snapshots.

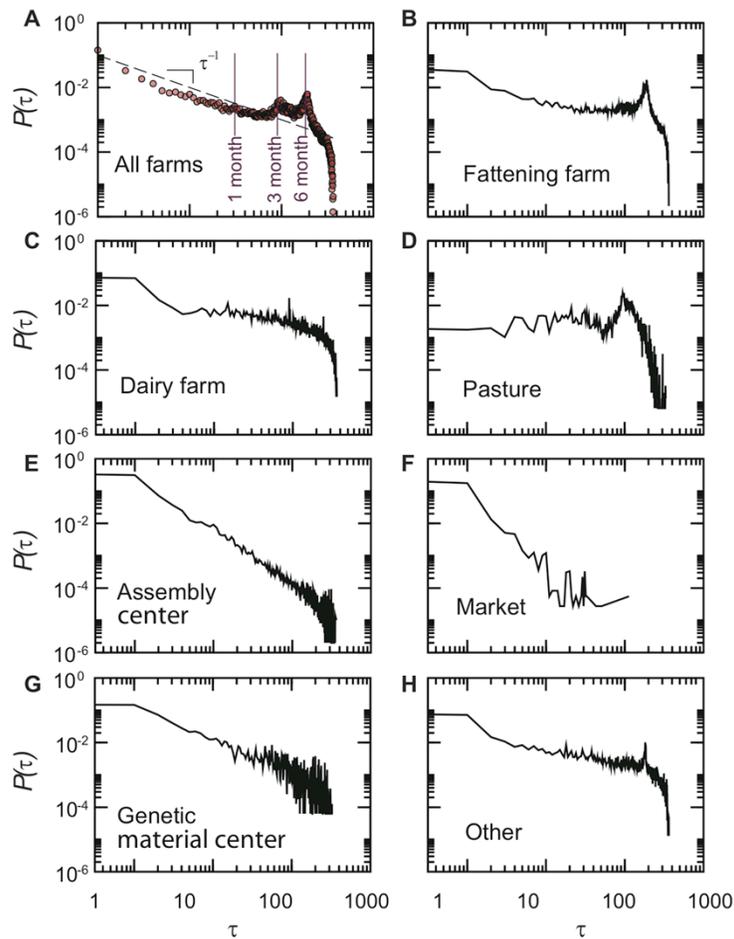

**Figure 4. Probability distributions of the time interval $\tau$ between two consecutive displacements of a bovine.** $\tau$ corresponds to the time during which the bovine stays at given premises. The seasonality behavior of breeding is clearly shown by the peaks at 3 and 6 months, while at shorter times the distribution behaves as $\tau^{-1}$. The global distribution is a convolution of the time distributions obtained for different farm types, shown in panels B to H.

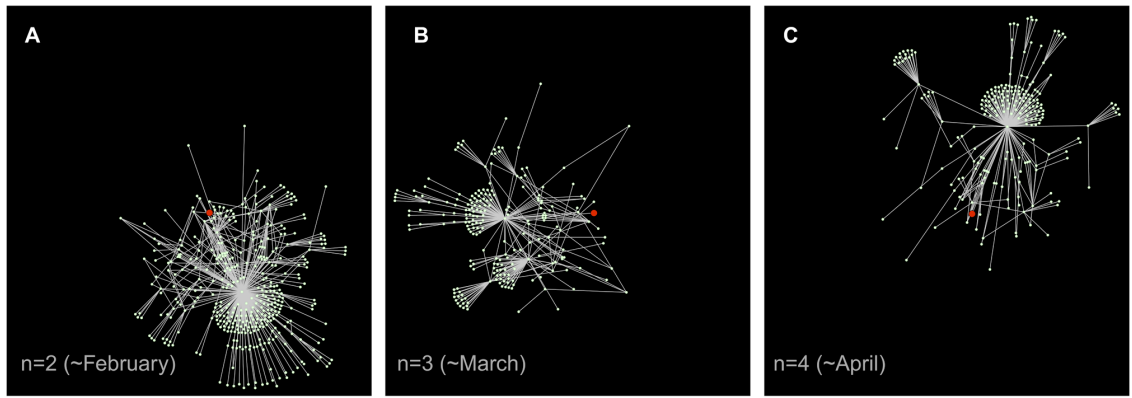

**Figure 5. Neighborhoods of a selected node in three consecutive monthly networks.** The subgraphs are obtained by showing all nodes within distance 3 from a selected node (in red in the figure), for consecutive monthly snapshots. The visualization highlights how the neighborhood of a given node may strongly change its structure in time. It is important to note that nodes that disappear from the plots may still be present in the network, but are not shown as they may be at distance larger than 3 from the seed, thus not belonging to its neighborhood.

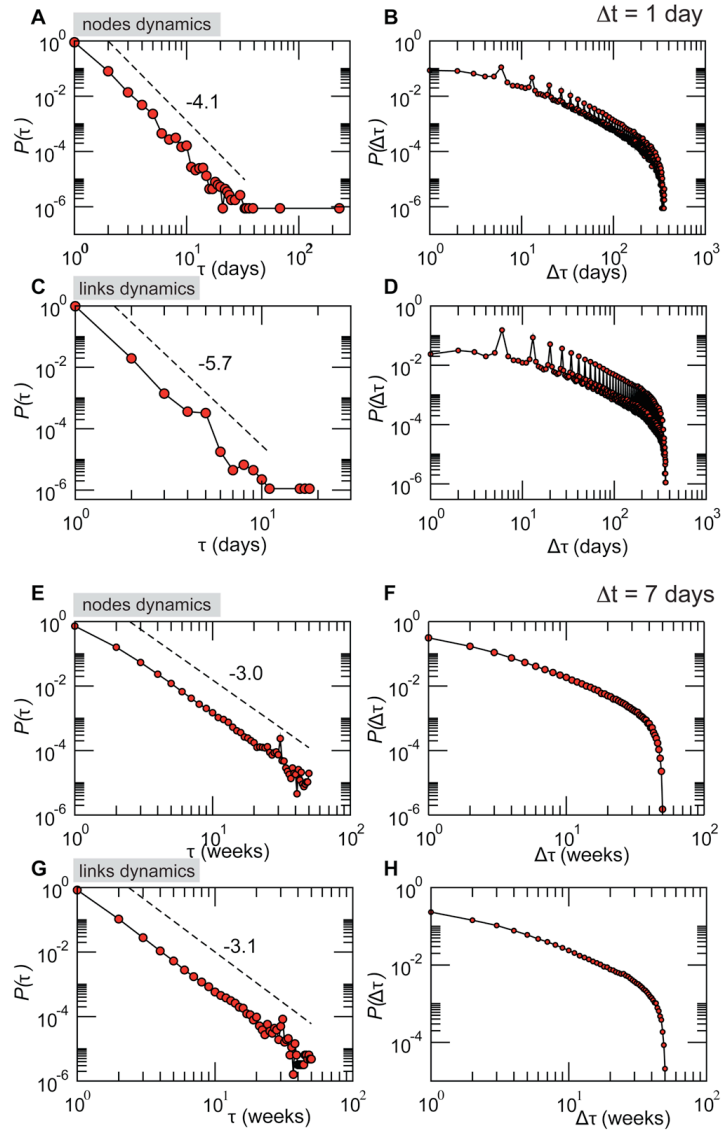

**Figure 6. Probability distributions of the duration $\tau$ of activity and of the duration $\Delta\tau$ of inactivity of nodes and links.** Results are reported for daily (panels A to D) and weekly (panels E to H) networks. In the daily case, weekend breaks are neglected as they are characterized by a much lower activity and clear weekly patterns (see Figure S3). The observed peaks in $P(\Delta\tau)$ of the daily networks correspond to inactivity periods of multiples of a week.

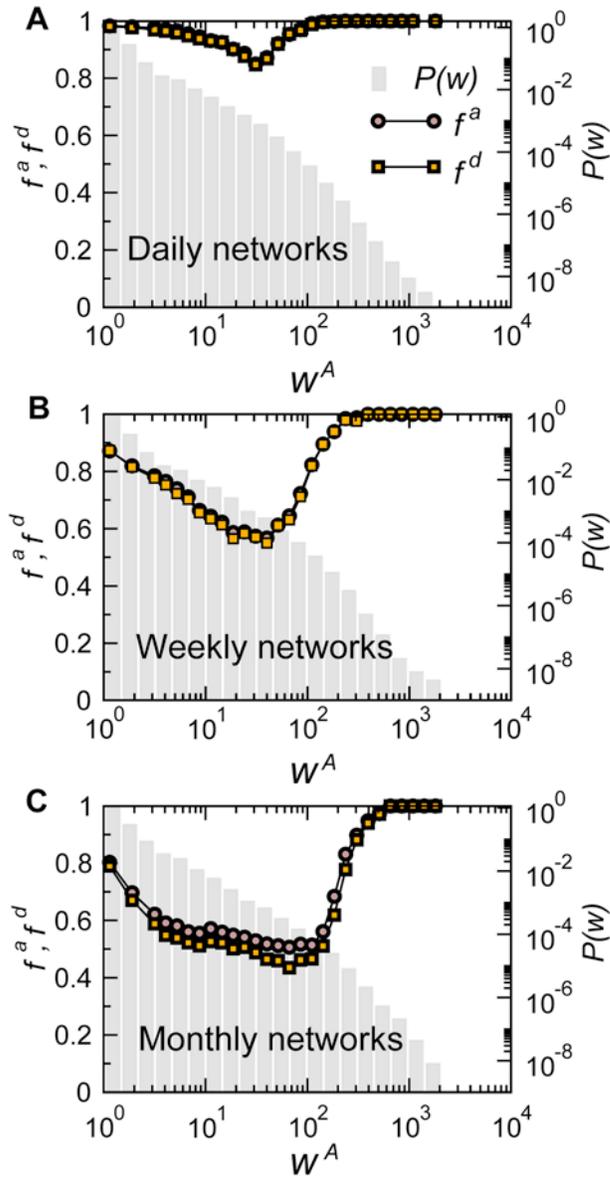

**Figure 7. Fraction of appearing/disappearing links as a function of the weight associated to the link.** The weight considered here counts the number of animals, $w^A$. Results for daily, weekly, and monthly networks are shown (panels A, B, C, respectively). As a reference, the weight distribution is also shown with a grey histogram.

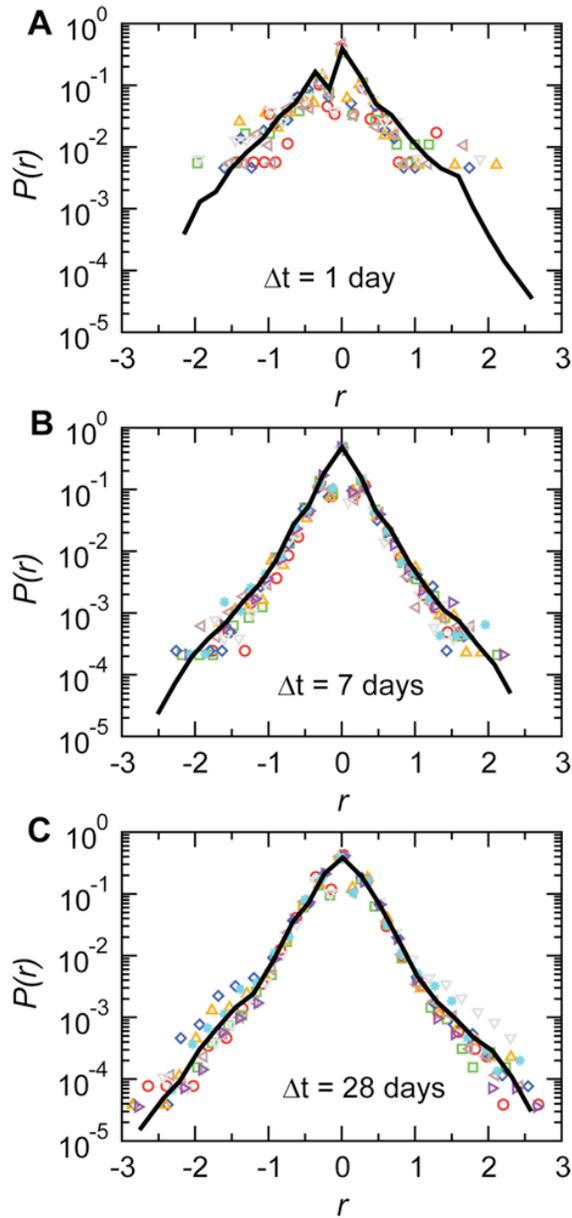

**Figure 8. Distributions of the growth rates of the number of bovines $w^A$ displaced along a connection.** The solid line represents the distribution of the growth rates considering all networks of a given aggregating time window $\Delta t$. Symbols corresponds a selection of snapshots.

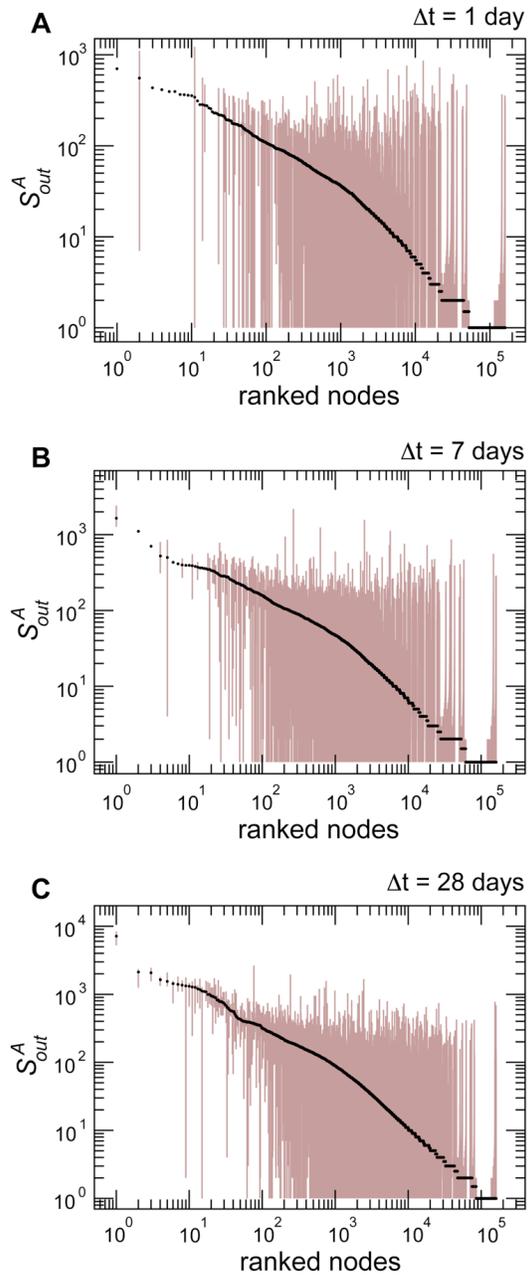

**Figure 9. Fluctuations of the total outgoing traffic of bovines of a given holding for various aggregating time windows.** The plot shows, for each holding of the system, the fluctuations of the values of $s_{out}^A$ assumed by each node during all snapshots of the $\Delta t$ under study. The median (black dots) and the 95% confidence interval (brown shaded area) of outgoing traffic are shown.

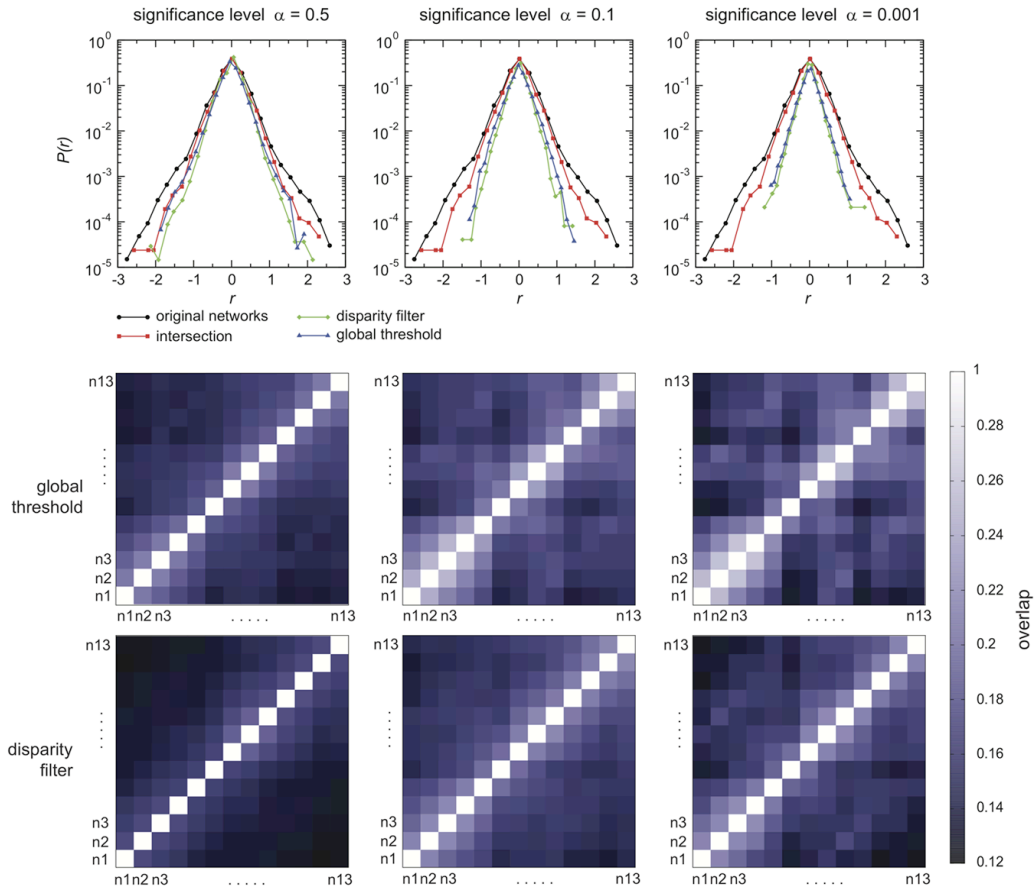

**Figure 10. Evolution of monthly network backbones.** Top: Distributions of the growth rates of the weights $w^A$ of the backbone links, where the network backbone is obtained under different filtering procedures. In each case, growth rates $r$ are measured only for links that are present in two successive backbones. Center and Bottom: Overlap between the backbones of monthly networks. The overlap measures the number of links common to the pair of networks under consideration, normalized by their total number of links. Backbones are obtained either with a global threshold filter (center row) or using a disparity filter (bottom row). Three values of the significance parameter $\alpha$ are considered.

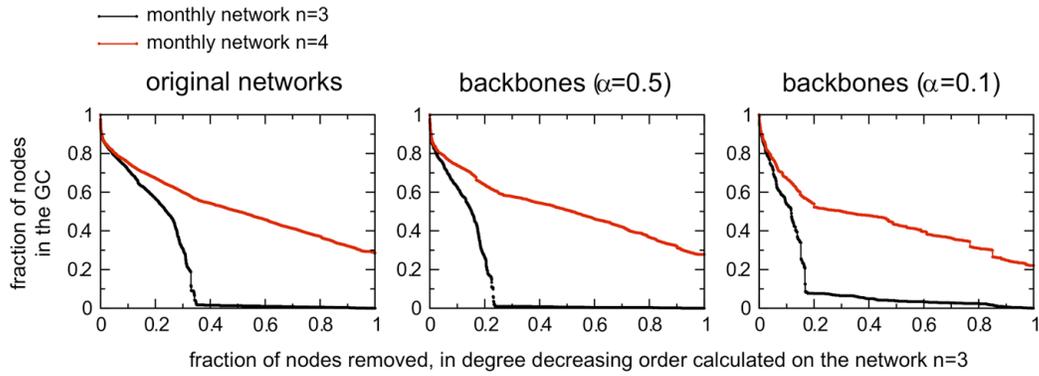

**Figure 11. Percolation analysis on consecutive monthly networks.** Two consecutive monthly snapshots ($n = 3$ and $n = 4$) have been considered. A list of nodes with decreasing degree is calculated on the snapshot $n = 3$, and is applied as a removal strategy for both networks. The same procedure has been performed on the corresponding network backbones obtained for two values of the significance parameter $\alpha$.

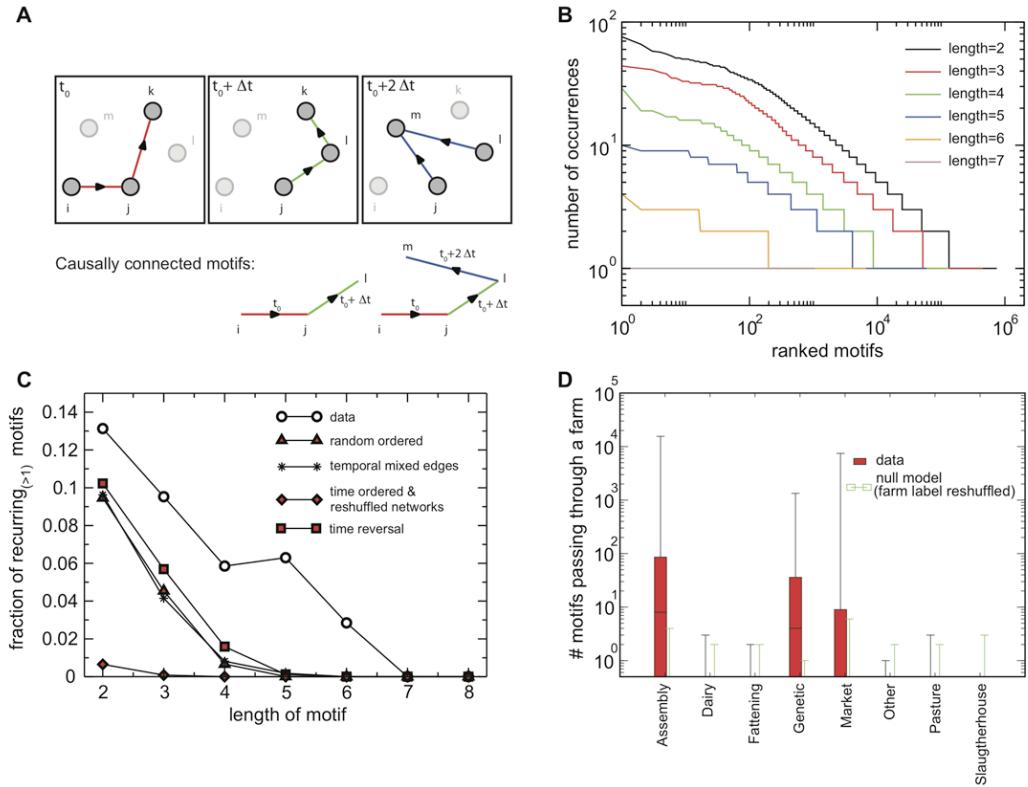

**Figure 12. Motifs: schematic representation and their occurrence.** A schematic example of the dynamics of a subset of the mobility networks is shown in panel A through three successive snapshots. The connections are color-coded according to the time at which they are active. A temporal motif is a temporal sequence of links such that the destination node of a link at time $t_0$ is the origin of another link at time $t_0 + \Delta t$. Two examples of motifs, of respective lengths 2 and 3, are shown below. We restrict the present study to the case of $\Delta t = 1$ day. Panel B shows the results on the presence of motifs, analyzed by counting the number of occurrences during the timeframe under study. The longer the motifs, the smaller the number of times they appear. By focusing only on the set of motifs that occur at least twice, panel C compares the size of this set (expressed as a fraction of the total) obtained from the empirical dataset with the sizes obtained through various randomization procedures (see main text). The results are shown as functions of the motifs length. In panel D the median and confidence intervals of the number of motifs passing through a farm depending on the farms type are shown,

together with the same computation for a null model in which the farm types are reshuffled at random.

## Supplementary Information

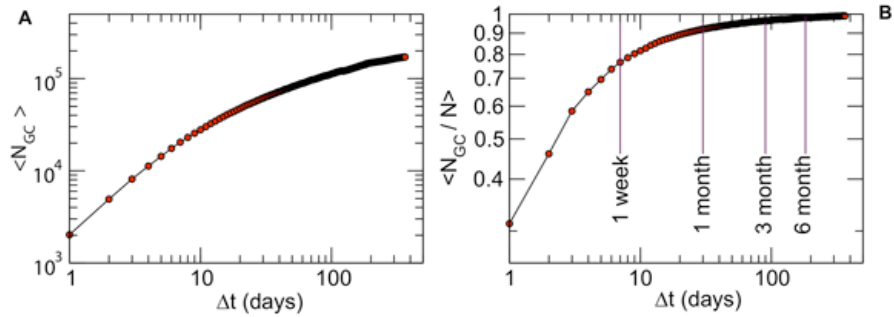

**Figure S1**. **Size of the giant component of the network for increasing aggregation timescale $\Delta t$.** Average number of nodes (panel A) and relative fraction with respect to the system size (panel B) of the giant component of networks aggregated on time windows of length $\Delta t$. As $\Delta t$ increases, the networks become more globally connected. Sizes are averaged over all snapshots obtained with a given value of $\Delta t$.

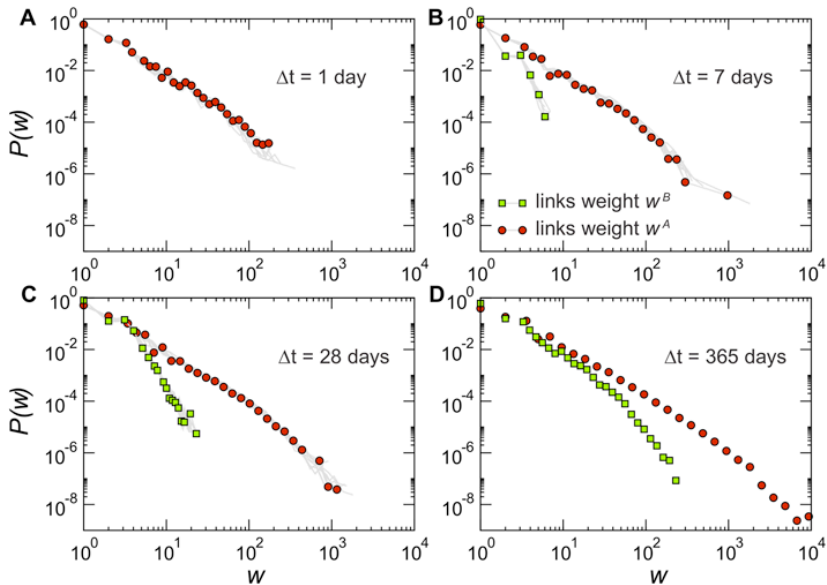

**Figure S2. Weight distributions for networks aggregated on different timescales $\Delta t$.** Red circles refer to the binned distributions of the weight $w_{ij}^A$, measuring the number of animals moved along the link *ij*, whereas green squares refer to the binned distributions of the weight $w_{ij}^B$ that counts the number of batches displaced along the link. The same representation of Figure 1 of the main text is adopted, with symbols representing the result of a particular snapshot, and grey lines the results obtained for a subset of the other snapshots. The cut-off of the $w_{ij}^B$ distributions is naturally fixed by the choice of the aggregating period $\Delta t$. The distribution of $w_{ij}^B$ for the daily networks has been omitted, since it is equal to 1 for $w_{ij}^B = 1$ and 0 elsewhere.

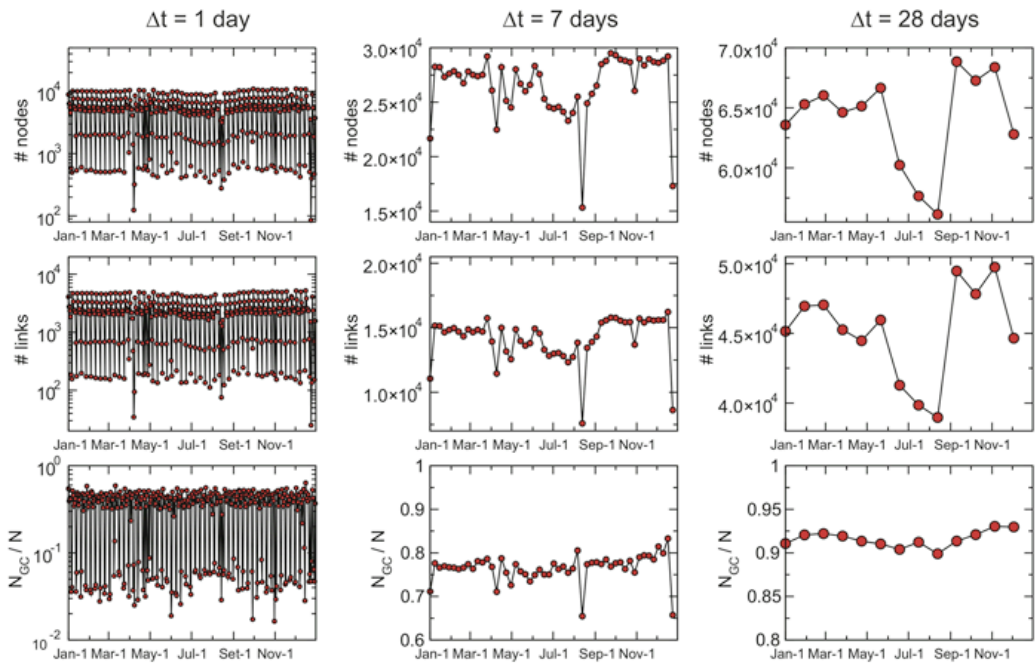

**Figure S3. Time evolution of the global static features of networks on different timescales.** The timeline of the number of nodes (top), the number of links (center), and the fraction of nodes in the giant component (bottom) are shown for daily, weekly, and monthly networks. Clear weekly and seasonal patterns are detected.

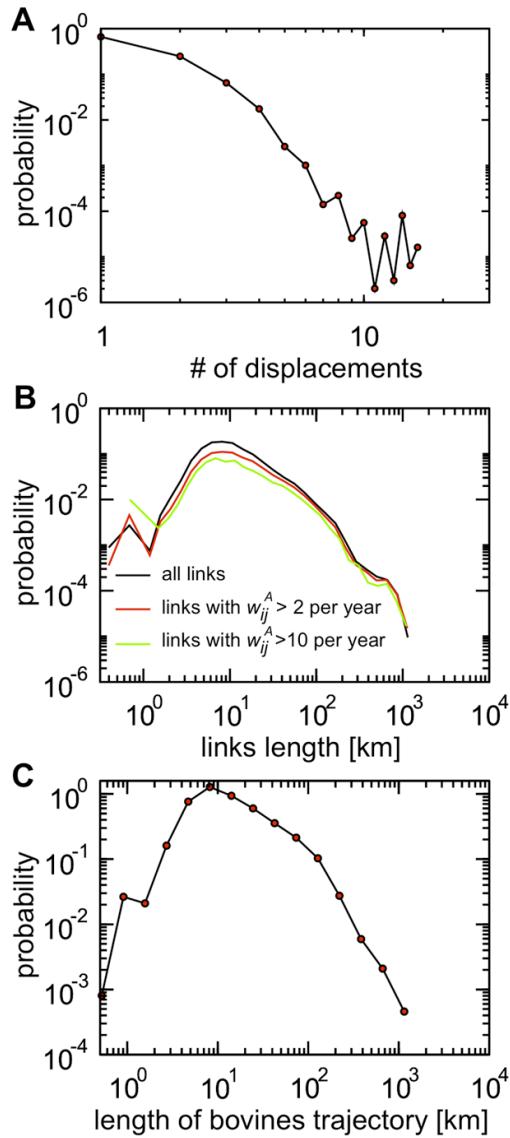

**Figure S4. Bovine activity.** Panel A shows the probability distribution of the number of displacements that a bovine experiences during one year. Panel B displays the probability distributions of the distances covered during a single displacement. Since many links correspond to the displacement of very few animals, the same distribution is shown with different thresholds, i.e. considering only links with at least 2 or 10 bovines displaced during the year under study. This corresponds to keeping respectively 42% and 13% of the original links. Panel C shows the probability distribution of the distances covered by a single animal during its trajectory in one year.

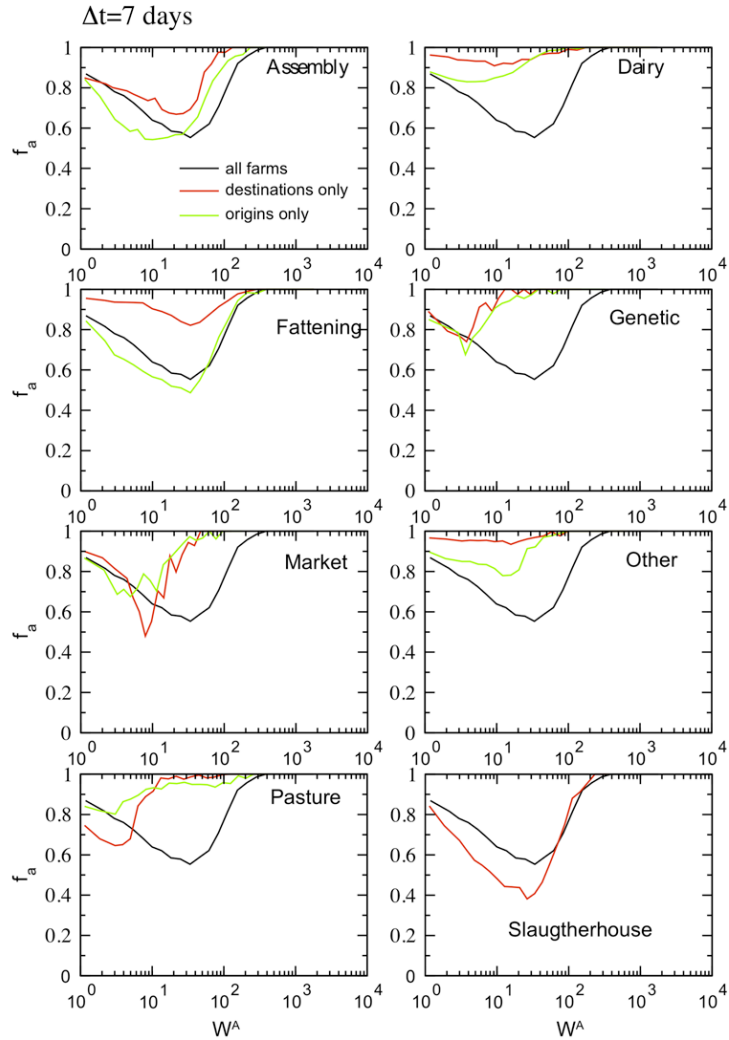

**Figure S5. Appearance of nodes by premises type.** The fraction of appearing links as a function of the weight $w^A$ associated to the link. The black curve refers to the total fraction of appearing links, the red curves are obtained considering only the links pointing to a given premises type, the green curves are obtained using only the links originating at a given premises type. The links' behavior depends on the nature of the displacement, i.e. the premises type of origin/destination. Consistent results are obtained for different timescales $\Delta t$. The fraction of disappearing links (not shown) displays an almost identical behavior (similarly to the results presented in Figure 7 of the main text).